\documentclass[aip,twocolumn,apl,graphicx]{revtex4}
\usepackage{graphicx}
\usepackage{dcolumn}
\usepackage{bm}

\begin{document}

\title{Ultra-short suspended single-wall carbon nanotube transistors} 

\author{J.~O. Island, V. Tayari, S. Yi\u{g}en, A.~C. McRae, A.~R. Champagne}
\email[]{a.champagne@concordia.ca}
\affiliation{Department of Physics, Concordia University, Montreal, Quebec, H4B 1R6 Canada}

\date{\today}

\begin{abstract}
We describe a method to fabricate clean suspended single-wall carbon nanotube (SWCNT) transistors hosting a single quantum dot ranging in length from a few 10s of nm down to $\approx$ 3 nm. We first align narrow gold bow-tie junctions on top of individual SWCNTs and suspend the devices. We then use a feedback-controlled electromigration to break the gold junctions and expose nm-sized sections of SWCNTs. We measure electron transport in these devices at low temperature and show that they form clean and tunable single-electron transistors. These ultra-short suspended transistors offer the prospect of studying THz oscillators with strong electron-vibron coupling.
\end{abstract}


\maketitle 

\begin{figure}
\includegraphics [width=3.35in]{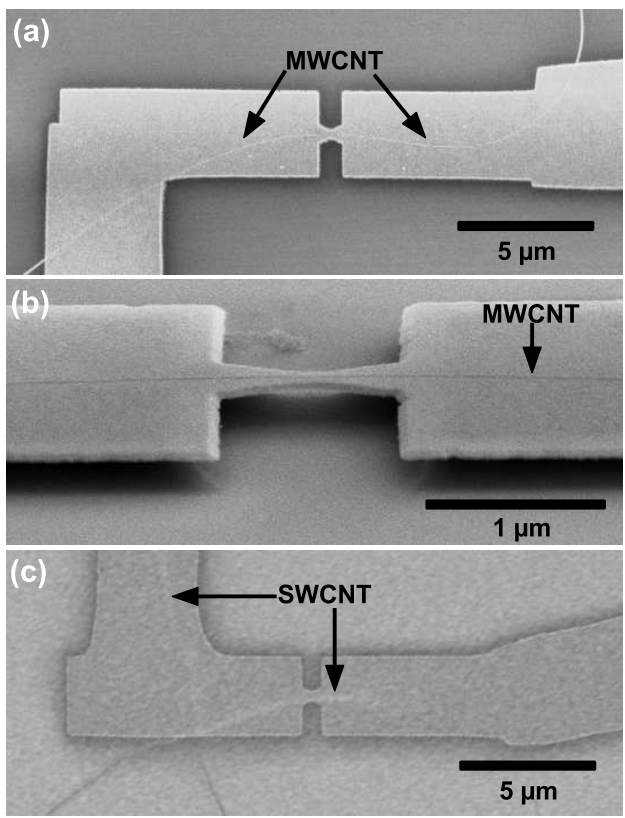}
\caption{\label{} (Color online.) SEM images of a MWCNT and a SWCNT breakjunction devices before electromigration. (a) A gold breakjunction directly on top of a MWCNT. The width of the junction is about 300 nm. (b) Tilted SEM image of the device in (a) after suspension. The MWCNT is clearly visible under the gold film. (c) A breakjunction aligned on top of a SWCNT (Device A). The image is obtained from the superposition of two SEM images, before and after the breakjunction is exposed, as a SWCNT is not visible under the gold film.}
\end{figure}

Much effort is currently focused on creating nanometer scale single-molecule electronic devices to explore the physics of interacting electrons and vibrons \cite{Park00,Pasupathy05, Paaske05, Sapmaz06, Parks07,Mariani09,Leturcq09,Lassagne09,Steele09,Cavaliere10, Ward11}, and to develop extremely small transistors \cite{Park00, Javey04, Champagne05, Ihn10} and electromechanical systems (NEMS)\cite{Bunch07,Huttel09}. One of the main hurdles to studying electron transport in single-molecule transistors is the poor control of the orientation and conformation of a molecule between electrodes. Single-wall carbon nanotubes (SWCNTs) are defect-free one-dimensional crystals which can be precisely aligned with respect to electrodes. Moreover, they are exceptionally strong and their charge carriers' mean-free path is $\gtrsim$ 10 nm even under larger bias voltage \cite{Javey04,Biercuk08}. Unfortunately, widely available nanofabrication methods cannot reliably fabricate clean suspended SWCNT devices shorter than a few 10s of nm. We note that a shadow evaporation method previously produced $\gtrsim$ 10 nm on-substrate SWCNT devices \cite{Javey04}. We report a controlled procedure to fabricate suspended SWCNT transistors hosting a single quantum dot (QD) as short as $\approx$ 3 nm. The suspension of these devices makes it possible to study them as NEMS, and to remove any contamination from the tubes by self-heating. We produced 9 suspended SWCNT transistors using this method, and measured 4 of them in detail. We show that they form clean and tunable QD transistors.

 To fabricate gold junctions on top of SWCNTs, we start with heavily-doped Si wafers with a 300 nm-thick SiO$_{2}$ film on their top side. The Si substrate will be used as a gate electrode. We define micron-sized Fe catalyst pads (5-6 ${\AA}$ thick) to seed SWCNTs, which we grow by chemical vapor deposition \cite{Jin07}. Using AFM we measure the diameter of our tubes to be $d$ = 1.2 $\pm$ 0.4 nm. We then expose micron sized gold alignment marks which we use to locate the nanotubes and align the gold junctions on them. To rapidly locate SWCNTs we briefly image the wafers in a SEM with a low accelerating voltage and magnification. This does not result in any noticeable contamination. Detailed SEM imaging of the devices, as shown in Fig.\ 1, is done after the gold film covers the nanotubes and protects them from contamination, or after all transport measurements are completed.  We use e-beam lithography to expose 40-nm thick gold wires (no adhesion layer) with a bow-tie shape on top of selected SWCNTs. The bow-tie junctions are one micron long and approximately 300 nm wide at their center (Fig.\ 1(a)). We use a wet BOE etch to suspend the bow-tie junction, Fig.\ 1(b). Note that the nanotube under the Au bridge is clearly visible in Fig.\ 1(b) because we used a MWCNT to demonstrate the procedure. Figure 1(c) shows the superposition of an SEM image taken before and after e-beam lithography on top of a SWCNT (Device A).

\begin{figure}
\includegraphics[width=3.35in]{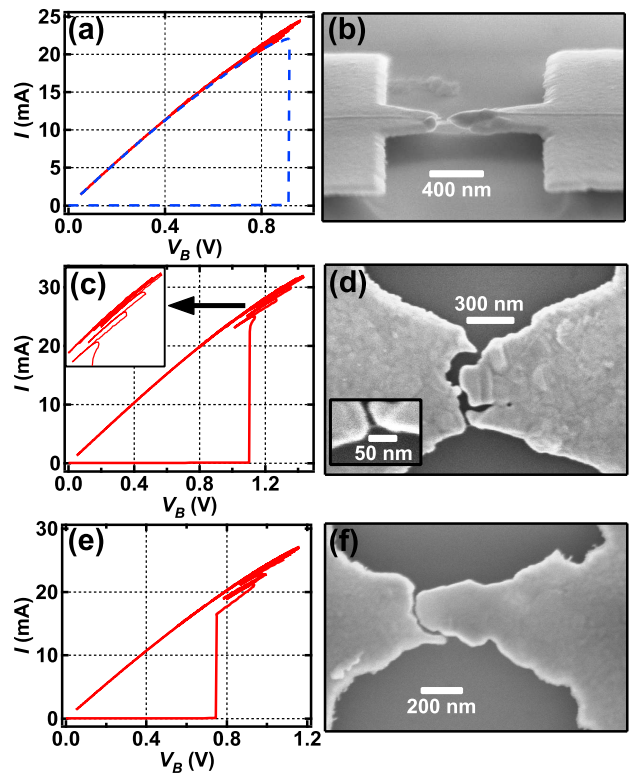}
\caption{\label{}(Color online.) Nanoetching technique. (a) $I-V_{B}$ characteristics during the electromigration of the device shown in Fig.\ 1(b). $V_{B}$ is ramped up until the feedback detects an increase in the resistance and rapidly lowers the bias. This procedure is repeated to gradually narrow the gold bridge (full red curve). The process can be stopped at any position in the $I-V_{B}$, and then the partially etched junction can be broken with a continuous voltage ramp (dashed blue curve). (b) Tilted SEM image of the 70 nm-long MWCNT device electromigrated in (a). (c) Nanoetching $I-V_{B}$ curve for SWCNT-Device A (Fig.\ 1(c)). The inset is a zoom on the gradual etching of the device. (d) Top view SEM image of Device A after breaking. The inset shows a zoom-in on the 22$\pm$5 nm-long SWCNT. The location of the tube corresponds to its position after e-beam lithography. (e) Electromigration curve for SWCNT-Device B. (f) SEM of Device B after breaking, showing a gap $\lesssim$ 10 nm.}
\end{figure}

The final fabrication step is to break the suspended gold bridges by electromigration \cite{Park99,Pasupathy05, Champagne05,Ward11} to uncover nm-sized sections of SWCNTs. We electromigrate our devices either in liquid Helium or in high-vacuum ($\leq 10^{-6}$ Torr) at $T \approx 4.2$K. We carefully control the rate of electromigration to tune the size of the gap and also avoid damaging the SWCNTs. Figure 2 (a) shows current-bias voltage, $I-V_{B}$, characteristics of this nanoetching process for the MWCNT device of Fig.\ 1(b). The resistance, $R=V_{B}/I$, is monitored in real time by a custom built feedback software which rapidly decreases $V_{B}$ when $R$ increases beyond a prescribed threshold $\Delta R$. The process can be repeated until the desired final resistance is reached. For instance, in Fig.\ 2(a) the software partially etched away the bridge in a first etching run made of many etching steps (red full curve, and inset of (c)), and then ramped down $V_{B}$ to zero. In a second run (blue dashed curve), we ramped up $V_{B}$ continuously (no feedback) to completely break the partially etched gold bridge. The resulting gap is shown in Fig.\ 2(b), where a 70-nm long section of the MWCNT is clearly visible. To demonstrate the procedure, this device was broken at a relatively high bias (0.9 V) and while ramping $V_{B}$ without feedback, which resulted in a large gap. The MWCNT was not damaged during this electromigration process, however SWCNTs are typically destroyed (broken) or damaged (very large charging energies or many QDs in series) when broken without feedback at relatively high $V_{B}$. Electromigration of gold wires takes place due to the forces exerted by both the electric field and the electron wind \cite{Park99}. The temperature at which gold wires break is only of a few hundred Kelvin \cite{Esen05}, much below the melting temperature of SWCNTs in high-vacuum. We believe that any damage to a SWCNT arises from the extremely high electric field across the shortest exposed section of tube right after the gold wire breaks. To minimize this field we could use narrower wires, but it would make the lithographic alignment difficult. We therefore break SWCNT devices either at lower bias, by using a two step process as in Fig.\ 2(a), or by breaking the bridge at higher $V_{B}$ while in the feedback mode by using a large $\Delta R$. This last method guarantees that the voltage is ramped down rapidly as soon as the wire starts breaking, to avoid applying a large $V_{B}$ across the bare tube, and gives small gaps. We used this approach to break Devices A and B respectively shown in Fig.\ 2(c)-(d), and (e)-(f). The inset of Fig.\ 2(d) shows a zoom-in on the lower portion of the breakjunction of Device A, where we can resolve the short SWCNT across the gap. This position matches precisely the location of the SWCNT under the bridge as determined from Fig.\ 1(c). The measured length of the tube is $L=$ 22$\pm$5 nm. Device B shown in panels (e)-(f), was broken at a lower $V_{B}$ and shows a smaller gap, $L\lesssim$ 10 nm. The exposed SWCNT cannot be resolve by SEM due to the proximity of the bright gold electrodes. Transport data presented below confirm that the SWCNT tube was not damaged during the electromigration. We can fabricate even shorter gaps by letting the electromigration software gently etch away the gold wire in successive etching steps until $R\sim h/e^2$. This creates atomic size gaps \cite{Strachan05}. However, these devices show considerable tunnel current and do not make well-behaved SWCNT transistors \cite{Guo04}.

\begin{figure}
\includegraphics[width=3.35in]{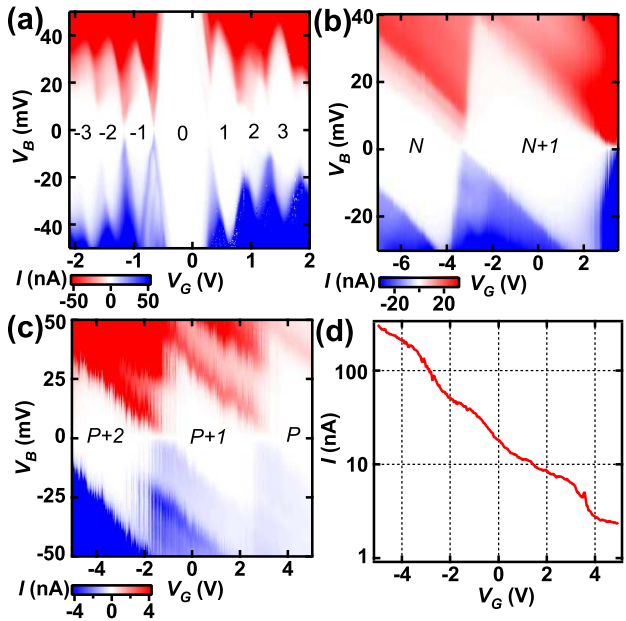}
\caption{\label{}(Color online.) Coulomb blockade transport data at $\approx$ 4.2K. (a), (b), and (c), $I-V_B-V_G$ for Device A, B and C respectively. The data show clean SWCNT QDs. The capacitances, $C_{G}$, extracted from the width of the diamonds correspond to tube lengths' of respectively $\approx$ 28, 3 and 4.5 nm. (d) $I$ vs. $V_{G}$ for Device C at $V_{B}=$ 120 mV, outside the Coulomb blockade region, showing a strong transistor effect.}
\end{figure}

We now characterize our devices with electron transport at $T\approx$ 4.2K. The majority of the 9 devices we studied were semiconducting, and formed QDs at low-temperature due to Schottky barriers at the Au/SWCNT interfaces. Figure 3(a) shows current-gate voltage-bias voltage ($I-V_{G}-V_{B}$) data for Device A. We see a very clear transistor effect with on and off states (Coulomb diamonds) \cite{vonDelft01}. The wide blockaded region around $V_{G}=0$ corresponds to the charge neutrality point where no carrier is present on the tube \cite{Biercuk08}. The fact that this depletion point is very close to zero gate voltage indicates that there is almost no residual doping and that any contamination adsorbed onto the tube during fabrication was ashen during the electromigration ($\sim$ 1 V across the suspended tube). The diamonds on the left and right represent the addition of holes and electrons on the dot. From the width of the diamonds we extract a gate capacitance $C_{G}=e/\Delta V_{G}$ = 0.31 aF \cite{vonDelft01}. In our devices, the gold bridge does not directly shield the SWCNT from the gate. We therefore expect that the capacitance per unit length of the tube can be roughly estimated using a wire over a plane model, $C/L = 2\pi\epsilon/(\cosh^{-1}(h/r))$ where $h$ is the wire to plane distance and $r$ the wire radius. We model $C_{G}$ as two such capacitors in series, respectively with vacuum and SiO$_{2}$ dielectrics. For most of our devices (Devices B and C below) the SWCNT was separated from the gate by 150 nm of vacuum and 150 nm of SiO$_2$. For Device A, $t_{oxide} =$ 100 nm and $t_{vac}\approx$ 25 nm due to BOE etching under the gold electrodes supporting the gold bridge. For Device A, we extract $L_{G}=$ 28$\pm$4 nm. This slightly overestimates the length compared to $L =$ 22$\pm$5 nm measured by SEM in Fig.\ 2(d), and confirms that the capacitance model gives a reasonable estimate of the length. Since $L_{G}\approx L$ for Device A, we learn that a single QD occupies the full length of the exposed SWCNT. Of the four devices we studied in detail, a second one was similar in length to Device A, with $L_{G}$ = 36$\pm$5 nm, and also showed clear Coulomb blockade diamonds signaling a single QD. We now focus on the shortest well-behaved transistor devices we made.

Figure 3(b) shows the transport data for the semiconducting SWCNT Device B (Fig.\ 2(e)-(f)). The data show one set of positive and negative threshold slopes typical of a single QD. The number of electrons on the dot can be tuned with $V_{G}$ from $N$ to $N+1$. We extract $C_{G}=$ 0.024 aF from the data, which is more than an order of magnitude smaller than $C_{G}$ for Device A, and gives a QD length $L_{G}\approx$ 3 nm. The data for Device B differs from Device A in two significant ways. The charge degeneracy point is not visible around $V_{G}=$ 0, and $I$ decreases from the right to left across the 2D plot. Both effects can be explained by the extremely short channel length of Device B. The Schottky barriers that form at the interface of metallic electrodes and the semiconducting SWCNT effectively transfer charge to the ends of the exposed tube (local gating). The length over which the nanotube's bands recover their normal dispersion in a heavily doped tube is of the order of a few times its diameter ($d\approx$ 1.2 nm) \cite{Odintsov00,Appenzeller02}. Thus, for Device B we expect that the channel length consists of two Schottky barriers each a few-nm long plus the QD dot which is $\approx$ 3 nm long. From the SEM image (Fig.\ 2(f)), the entire device is $\lesssim$ 10 nm, thus the QD is very close to the electrodes and heavily gated by them. This explains why the charge degeneracy point is not visible in Fig.\ 3(b) where the number of electrons can only be tuned by $\pm$ 1. The decrease of $I$ from the right to the left side comes from the semiconducting bandgap of the SWCNT. Close inspection of the data reveals additional lines running parallel above the Coulomb diamond labeled $N$. These lines represent excited states of the QD \cite{vonDelft01}, and intersect the $N+1$ diamond at $\Delta V_{B}\approx$ 5 mV. This energy scale is more than an order of magnitude too small to correspond to electronic excitations of the QD \cite{Biercuk08}. Rather it matches the energy of longitudinal vibronic modes \cite{Sapmaz06,Leturcq09}, $\Delta E_{v} = \hbar v_{ph} \Delta q$, where $v_{ph} \approx 2.4\times 10^4$ m/s  is the average group velocity and $\Delta q = \pi/L_{v}$. The extracted length of the oscillator is $L_{v}$ = 10$\pm$1 nm. This further supports the fact that the QD size is significantly smaller than 10 nm, since the oscillator's length corresponds to the entire length of the suspended tube which includes the two tunneling (Schottky) barriers.

Devices B and C were made on the same nanotube by placing two junctions along its length. From SEM imaging, the gap length of Device C is also $\approx$ 10 $nm$. Figure 3 (c) shows data for Device C which is p-doped, i.e. $V_{G}$ tunes the number of holes $P$, rather than n-doped as Device B. We extract $C_{G}=$ 0.037 aF which corresponds to a QD size of $L_{G}\approx$ 4.5 nm. A similar change from n to p doping when $L_{G}$ increases, and is comparable to the Schottky barrier's thickness, was observed previously \cite{Park01}. The conductivity of Device C is a few times lower than for Device B confirming that its tunnel barriers are thicker. Figure 3(d) shows an $I-V_{G}$ curve for Device C, but with $V_{B}=$ 120 mV to probe the device outside of the Coulomb blockaded window. The data shows an excellent transistor effect with $I_{on}/I_{off} > 100$, and suggests that it should be possible to achieve a good transistor effect in these ultra-short devices up to room temperature.

We developed a method to controllably fabricate suspended SWCNT containing single QDs ranging in length from $\approx$ 3 nm up to a few 10s of nm. Even the shortest devices showed a clean transistor effect and we could tune the number of electrons or holes in their ground state. We plan to use these few-nm QDs to explore the strongly interacting elecron-vibron regime (e.g. Fig. 3(b)) \cite{Mariani09}, and develop THz-NEMS which can be studied by DC electron transport. This work was supported by NSERC, CFI (Canada), FQRNT (Quebec), and Concordia University. We made use of the QNI (Quebec Nano Infrastructure) cleanroom network.

\end{document}